\pdfoutput=1
\documentclass[a4,prb,twocolumn,amsfonts,amssymb,showpacs]{revtex4-1}
\usepackage{amsmath}
\usepackage{graphicx}
\usepackage{color}
\usepackage{hyperref}
\usepackage{psfrag}
\usepackage{stmaryrd}
\usepackage{amssymb}
\usepackage{wasysym}

\begin{document}

\title{Excitation of inertial modes in a closed grid turbulence experiment under rotation}

\author{Cyril Lamriben}
\author{Pierre-Philippe Cortet}
\email{ppcortet@fast.u-psud.fr}
\author{Fr\'{e}d\'{e}ric Moisy}
\affiliation{Laboratoire FAST, CNRS UMR 7608, Universit\'{e}
Paris-Sud, Universit\'{e} Pierre-et-Marie-Curie, B\^{a}t. 502,
Campus universitaire, 91405 Orsay, France}
\author{Leo R. M. Maas}
\affiliation{Royal Netherlands Institute for Sea Research, Texel,
The Netherlands}

\date{\today}

\begin{abstract}
We report an experimental study of the decay of grid-generated
turbulence in a confined geometry submitted to a global rotation.
Turbulence is generated by rapidly towing a grid in a
parallelepipedic water tank. The velocity fields of a large number
of independent decays are measured in a vertical plane parallel to
the rotation axis using a corotating Particle Image Velocimetry
system. We first show that, when a ``simple'' grid is used, a
significant amount of the kinetic energy (typically 50\%) is
stored in a reproducible flow composed of resonant inertial
modes. The spatial structure of those inertial modes, extracted by
band-pass filtering, is found compatible with the numerical
results of Maas [Fluid Dyn. Res. \textbf{33}, 373 (2003)]. The
possible coupling between these modes and turbulence suggests that
turbulence cannot be considered as freely decaying in this
configuration. Finally, we demonstrate that these inertial modes
may be significantly reduced (down to 15\% of the total energy) by
adding a set of inner tanks attached to the grid. This suggests
that it is possible to produce an effectively freely decaying
rotating turbulence in a confined geometry.
\end{abstract}

\maketitle

\section{Introduction}
\label{sec:intro}

Translating a grid in a closed volume of fluid is a standard way
to generate an approximately homogeneous and isotropic turbulence
in order to investigate its temporal decay. Although the level of
homogeneity and isotropy of the turbulence generated in this way
is not as good as in the more conventional configuration of a
fixed grid in an open wind tunnel,\cite{cbc1966,mohamed1990} this
closed flow configuration has proved to be very useful when a
compact system is needed, and in particular when experiments are
performed in a rotating
frame.\cite{ibbetson1975,dalziel1992,morize2005,bewley2007,staplehurst2008}
Apart from early experiments performed in wind tunnels with a
rotating section,\cite{wigeland,jacquin1990} all rotating
grid-generated turbulence experiments since then are based on
oscillated or translated grids in closed containers.

Although grid turbulence in open and closed geometries share
similar properties, large scale recirculations are more likely
produced in the closed configuration.\cite{mckenna2004} These
reproducible flows, which will be called mean flows in the sequel,
can be easily extracted thanks to ensemble average over several
independent realizations. They may actually be triggered from
inhomogeneities in the wake of the grid, or from spontaneous
symmetry breaking in a perfectly symmetric geometry. In the case
of decaying turbulence, although the kinetic energy of the mean
(i.e., ensemble-averaged) flow may be negligible compared to that
of turbulence soon after the grid translation, it may persist over
large times and possibly influence the final period of decay of
turbulence.

In the presence of background rotation, grid-generated turbulence
may excite inertial waves. These waves are anisotropic transverse
dispersive waves, which propagate through a rotating fluid because
of the restoring nature of the Coriolis
force.\cite{greenspan,cortet2010} Their pulsation $\sigma$ lies in
the range $[0, 2 \Omega]$, where $\Omega$ is the rotation rate,
indicating that these waves can be excited when the characteristic
time of the considered flow is of order of the rotation period,
i.e. when the Rossby number is of order unity. These inertial
waves may originate either from a mean flow, in which case they
are reproducible and can be detected in the ensemble-average, or
from turbulence itself, in which case they are non reproducible
and cancel out by phase mixing in the ensemble-average.

In grid-generated turbulence experiments performed in a closed
container, still in the presence of  rotation, reproducible
oscillations induced by inertial modes have been
reported.\cite{dalziel1992,bewley2007,moisy2010} A first mechanism
responsible for these modes is the resonance of multiply reflected
inertial waves over the
container.\cite{fultz1959,phillips1963,mcewan1970} Another
possibility is that they originate from direct conversion of the
energy of the large scale recirculations initiated by the grid
translation, which were also present in the non-rotating case,
when the characteristic time of these recirculations decreases
down to the order of the rotation period.

Inertial modes can be found in general in a container when the
walls are either normal or parallel to the rotation axis, whereas
wave attractors are found when sloping walls are
present.\cite{manders2003} Only the former are considered in the
following. Their resonant frequencies can be derived analytically
only in some specific geometries, such as the so-called Kelvin
modes in a cylinder rotating about its symmetry
axis.\cite{batchelor1967} In the case of a parallelepipedic
container, the frequencies and the spatial structure of the
eigenmodes have been characterized in detail  by
Maas\cite{maas2003} for an inviscid fluid from numerical
simulations.

Inertial modes excited in grid-generated turbulence have been
first observed  in a parallelepipedic channel with a free surface
by Dalziel,\cite{dalziel1992} who pointed their potential
influence on the decay of the turbulent component of the flow.
Inertial oscillations are also clearly visible in the experiments
of Morize and Moisy\cite{morize2006} (with a rigid upper surface)
and Moisy \textit{et al.}\cite{moisy2010} (with a free surface).
They have been characterized by Bewley \textit{et
al.}\cite{bewley2007} in a set of two experiments, in which a grid
is rapidly drawn in liquid helium or nitrogen in cylindrical and
squared geometries. These authors found good agreement between the
measured frequencies and the numerical results of
Maas\cite{maas2003} for various aspect ratios. They conclude that
translating a grid in a closed rotating container does not allow
to generate freely decaying turbulence, because a significant
amount of the energy goes into inertial modes. In all these
experiments, these inertial modes are detected in the
ensemble-averaged flow, suggesting that they are more likely
produced by reproducible flow features, than from turbulence.

The aim of the present paper is to investigate in more details the
structure of the inertial modes excited when towing a grid in a
rotating square container, and to explore to what extent those
modes may be reduced. An experimental setup similar to that of
Morize \textit{et al.}\cite{morize2005,morize2006} has been
mounted on a new rotating platform, allowing for Particle Image
Velocimetry measurements in a vertical plane. We demonstrate that,
by attaching a set of inner sidewalls to the grid, the amount of
energy stored in the inertial modes may be drastically reduced. A
similar configuration has been first shown by Staplehurst
\textit{et al.}\cite{staplehurst2008} to reduce significantly the
large-scale recirculations in the absence of rotation. We
demonstrate here how a configuration inspired from their work
inhibits the excitation of inertial modes, by an enhanced
conversion of the mean flow energy into turbulence.

\section{Experimental setup}

\subsection{Tank and rotating platform}

\begin{figure}
\centerline{\includegraphics[width=7.5cm]{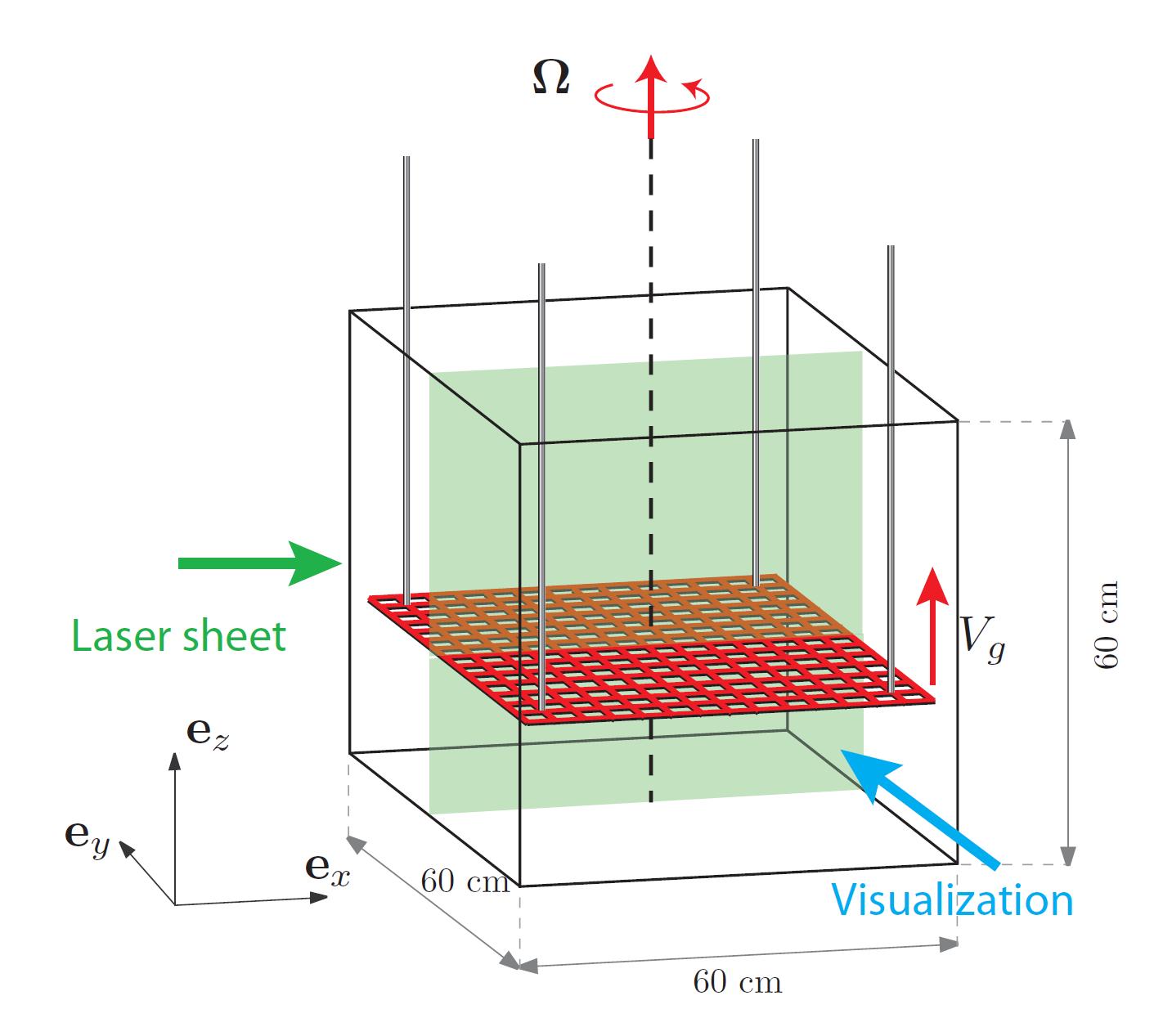}}
\caption{(Color online) Schematic view of the experimental setup,
in the ``simple grid'' configuration. The $60 \times 60 \times
60$~cm$^3$ tank is filled with 52~cm of water and is rotating at
an angular velocity of $\Omega=0.84$~rad~s$^{-1}$. A square grid
of 40~mm mesh is towed by four shafts from the bottom to the top
at constant velocity $V_g=0.70$~m~s$^{-1}$. PIV measurements in a
vertical plane ($x,z$) are achieved in the rotating frame, based
on a laser sheet illuminating the vertical plane ($x,z$) and a
camera aiming normally at it.} \label{fig:setup}
\end{figure}

Only the ``simple grid'' configuration, used in
Sec.~\ref{sec:sgc}, is described here. The modified configuration
with additional inner walls is detailed in Sec.~\ref{sec:giw}.

The experimental setup consists in a cubic glass tank, of lateral
side $L=60$~cm, filled with 52~cm of water (see
Fig.~\ref{fig:setup}) and mounted on a precision rotating
turntable of 2~m in diameter.\cite{cortet2010,gyroflow} The
angular velocity $\Omega$ of the turntable is set to
0.84~rad~s$^{-1}$ (8 rpm), with relative fluctuations $\Delta
\Omega / \Omega$ less than $5 \times 10^{-4}$. The rotation of the
fluid is set long before the experiment (at least half an hour) in
order for transient spin-up recirculations to be damped and
therefore to achieve a solid body rotation. A horizontal cover is
placed at a height of $H=49$~cm, defining the upper boundary of
the flow.

Turbulence ---and inertial modes--- is generated by rapidly towing
a square grid at a constant velocity $V_g=0.70$~m~s$^{-1}$ from
the bottom to the top of the tank. During the subsequent decay of
turbulence, the grid is kept fixed, at a height of 46~cm, slightly
below the cover.  The grid consists in 8~mm thick square bars with
a mesh size $M=40$~mm, and a ratio of solid to total area of 0.44.
It is rigidly attached at its four corners to a servo-controlled
brushless motor ensuring the vertical translation of the grid,
with acceleration and deceleration phases less than 18\% of the
total translation time. The Reynolds number based on the grid mesh
is $Re_g = V_g M / \nu = 28~000$, and the Rossby number is $Ro_g =
V_g / 2\Omega M = 10.4$, indicating that the flow in the close
wake of the grid is fully turbulent and weakly affected by the
rotation.

\subsection{PIV measurements}
\label{sec:piv}

Velocity fields in a vertical plane ($x,z$) are measured using a
2D Particle Image Velocimetry (PIV) system.\cite{DaVis} The flow
is seeded with 10~$\mu$m tracer particles, and illuminated by a
corotating vertical laser sheet passing through the center of the
tank and generated by a 140~mJ Nd:YAg pulsed laser. The entire
$60\times46$~cm$^{2}$ flow section is imaged through a transparent
side of the tank with a double-buffer high-resolution $2048\times
2048$~pixels camera, corotating with the tank and aiming normally
at the laser sheet. During the decay of turbulence, $428$ image
pairs are acquired at a sampling rate of 2~Hz. Since the typical
flow velocities decrease with time, the delay between the two
successive images of a pair is made to gradually increase during
the acquisition sequence, from 10 to $68$~ms, so that the typical
particles displacement remains constant, of order of 5 pixels,
during all the decay. PIV computations are then performed over
image pairs, on $32\times 32$ pixels interrogation windows with
$50\%$ overlap, leading to a spatial resolution of 4.9~mm.

\subsection{Reynolds decomposition}

\begin{figure}
\centerline{\includegraphics[width=7cm]{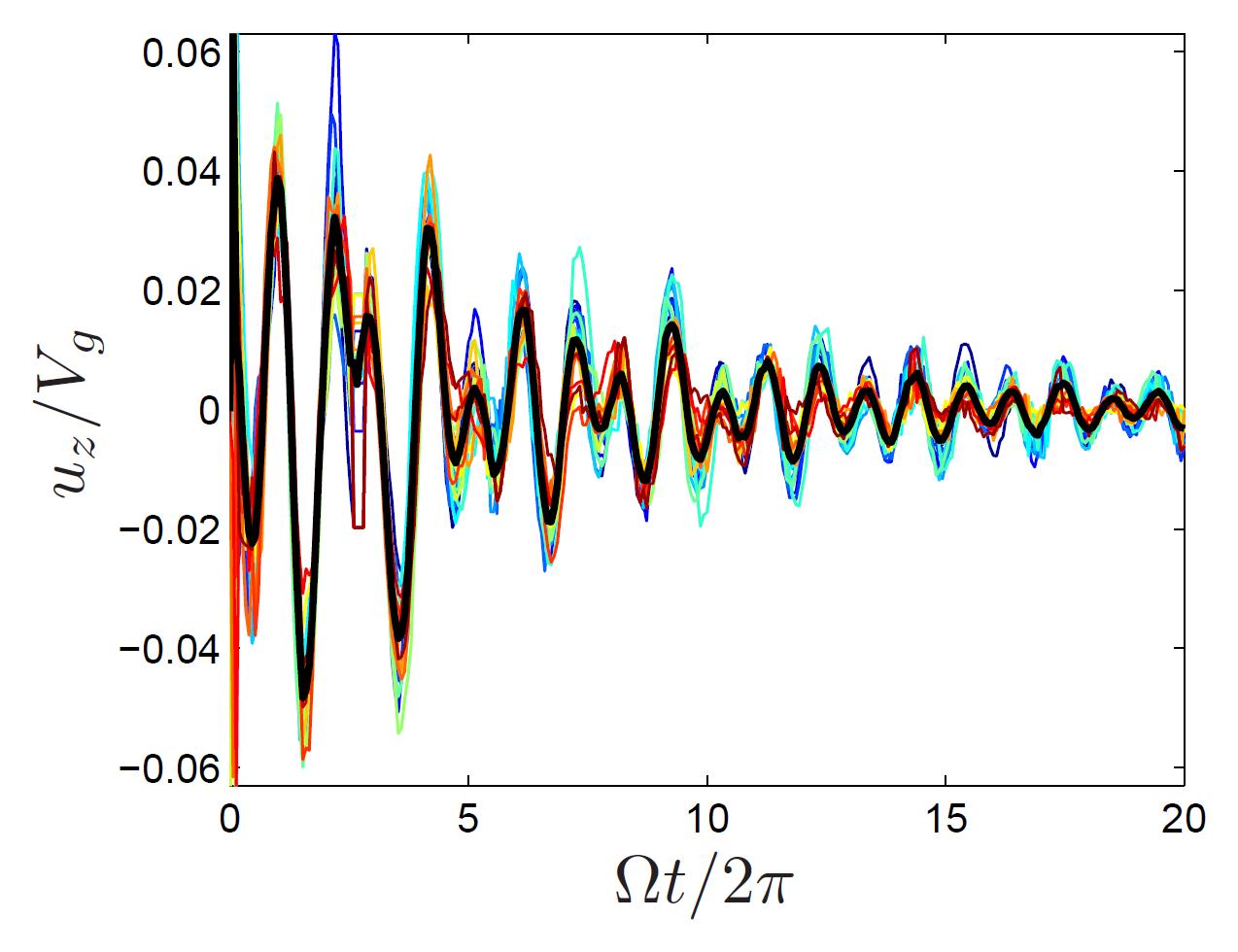}}
\caption{(Color online) Vertical velocity $u_z(x_0, z_0, t)$ at
the center of the flow for 20 realizations performed at
$\Omega=0.84$~rad~s$^{-1}$ (in various colors). We superimposed in
black thick line the ensemble average of these time series.}
\label{fig:vitcol}
\end{figure}

In Fig. \ref{fig:vitcol}, we show the time series of the vertical
velocity $u_z(x_0, z_0, t)$ at the center of the flow for 20
independent realizations of the decay.  The origin $t=0$ is
defined as the time at which the grid reaches the top of the tank.
The ensemble-average of those realizations is also shown in bold
line. This plot clearly illustrates that the flow consists in
well-defined oscillations excited by the grid, of characteristic
timescale of about one period of rotation, superimposed to
non-reproducible turbulent fluctuations. Those oscillations
actually correspond to inertial modes, and their amplitude is
clearly of the order of the turbulence.

In order to investigate properly the dynamics of the inertial
modes and of the turbulence, and the possible coupling between the
two, we introduce the standard Reynolds decomposition of the
velocity:
\begin{equation}
\label{eq:Redecomp}
\mathbf{u}(\mathbf{x},t)=\mathbf{U}(\mathbf{x},t)+\mathbf{u^\prime}(\mathbf{x},t).
\end{equation}
Here $\mathbf{u}(\mathbf{x},t)$ is the total velocity field,
$\mathbf{U}(\mathbf{x},t) \equiv
\overline{\mathbf{u}}(\mathbf{x},t)$ its ensemble average (i.e.
the reproducible component of the flow) and
$\mathbf{u^\prime}(\mathbf{x},t)$ its turbulent component. The
overbar $\overline{~\cdot~}$ stands for the ensemble average over
several independent realizations of the flow.  Note that, although
the ensemble average is often approximated by a temporal or a
spatial average in most turbulence experiments, the use of true
ensemble averages here is critical to separate properly the
reproducible non-stationary component of the flow from the
turbulence.

The Reynolds decomposition (\ref{eq:Redecomp}) naturally leads to
introduce three kinetic energies, characterizing the total, mean
and turbulent flows respectively, and defined as:
\begin{eqnarray}\label{eq:kturb}
k_{tot}(t) &=&  \langle \overline{\mathbf{u}^2} (\mathbf{x},t) \rangle, \nonumber \\
k_{mean}(t) &=& \langle \mathbf{U}^2(\mathbf{x},t) \rangle, \nonumber \\
k_{turb}(t) &=& \langle \overline{\mathbf{u^\prime}^2}
(\mathbf{x},t) \rangle, \label{eq:ke}
\end{eqnarray}
satisfying $k_{tot}(t)=k_{mean}(t)+k_{turb}(t)$. Here the brackets
$\langle \cdot \rangle$ denote the spatial average over the whole
fluid volume.

In practice, only the two velocity components $u_x$ and $u_z$ can
be measured, the measurements being restricted to a vertical plane
$(x,z)$ located at mid-width of the tank. The measured kinetic
energies must therefore be considered as approximations of the
true ones. For a statistically homogeneous and isotropic turbulent
component, one would simply have $k_{turb}^{mes} = 2 k_{turb} /
3$. The situation is more complicated for the energy of the mean
flow for two reasons: (i) during one period of a given inertial
mode, the kinetic energy is basically exchanged between the
measured and the non-measured velocity components; (ii) the ratio
between the energy averaged over the whole tank and the energy
averaged over the measurement plane $(x,z)$ only depends on the
details of the spatial structure of each mode, which are not known
a priori. For those reasons, we do not apply here any correcting
weight to the velocity components when computing kinetic energies,
and we simply define
\begin{equation}\label{eq:kturbexp}
k_{tot}(t) =  \langle \overline{{u_x}^2} \mathbf(x,z,t) +
\overline{{u_z}^2} \mathbf(x,z,t) \rangle_{x,z}
\end{equation}
(with $\langle \cdot \rangle_{x,z}$ the average over the vertical plane),
and similarly for $k_{mean}$ and $k_{turb}$. Note that those modified
definitions still satisfy the relation $k_{tot}(t)=k_{mean}(t)+k_{turb}(t)$.

\section{Inertial modes produced by the simple grid configuration}
\label{sec:sgc}

\subsection{Kinetic energy decay}

\begin{figure}
\centerline{\includegraphics[width=8cm]{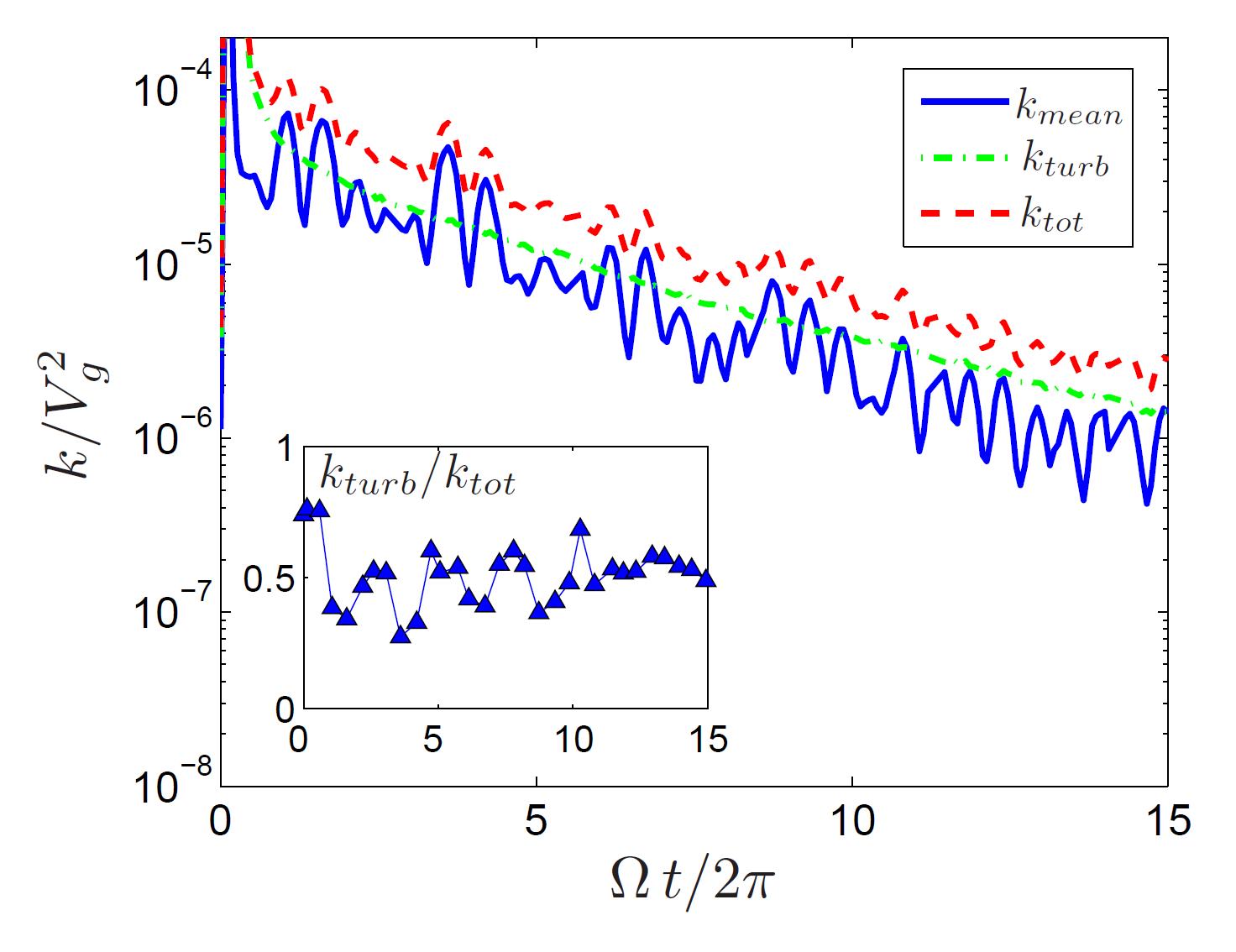}}
\caption{(Color online) Total (dashed), mean (continuous) and
turbulent (dashed-dotted) kinetic energies as a function of the
number of tank rotations $\Omega\,t/2\pi$ from 40 realizations
performed at $\Omega=0.84$~rad~s$^{-1}$. Inset: Ratio of turbulent
to total kinetic energy $\alpha$ (\ref{eq:alpha}), measured at
times $t_n$ of maximum mean energy.}\label{fig:decay1}
\end{figure}

In Fig. \ref{fig:decay1}, we present the time evolution of the
three kinetic energies (\ref{eq:ke}) computed from 40 decay
realizations performed at $\Omega=0.84$~rad~s$^{-1}$. Both the
energy of the total and the mean flow show, superimposed to their
overall decay, marked oscillations corresponding to the inertial
modes. On the other hand, the turbulent energy shows a monotonous
decrease, suggesting a good separation between the reproducible
and non-reproducible components of the flow. At this point, it is
important to note that the kinetic energy of the mean flow
$k_{mean}(t)$ is of the same order than the turbulent one
$k_{turb}(t)$.

The oscillations of $k_{tot}(t)$ and $k_{mean}(t)$ are only due to
the 2-dimensions and 2-components restriction of the PIV
measurements. Indeed, a monotonous decay should be expected for
the true energies (\ref{eq:ke}) computed from the 3 velocity
components averaged over the whole fluid volume. We  expect
therefore that the total and mean kinetic energies are correctly
estimated only at the maximum of their oscillations. In order to
evaluate the relative amount of turbulent and mean energy in the
total flow, we shall therefore consider only the times $t_n$ at
which $k_{mean}(t)$ is maximum, and we introduce the ratio:
\begin{equation}
\alpha(t_n)=\frac{k_{turb}(t_n)}{k_{tot}(t_n)}.
\label{eq:alpha}
\end{equation}
In the inset of Fig.~\ref{fig:decay1}, we actually see that
turbulence represents only $50 \pm 10\%$ of the energy in the
flow. This quite low ratio indicates that the turbulence produced
with this ``simple'' grid configuration is evenly distributed
among the reproducible inertial modes and the ``true turbulence''.
It is therefore questionable to consider this turbulence as freely
decaying. Indeed, the possible coupling between the turbulence and
the mean flow, which cannot be investigated at this point, may
prevent the turbulence to decay freely, as it is continuously fed
through energy transfer from the inertial modes.

\subsection{Fourier analysis of the inertial modes}

In order to characterize in more details the structure of the flow
generated by the grid translation, we perform a temporal Fourier
analysis of the ensemble-averaged flow $\mathbf{U}(\mathbf{x},t)
\equiv \overline{\mathbf{u}}(\mathbf{x},t)$. Assuming that this
mean flow is composed of inertial modes only,
$\mathbf{U}(\mathbf{x},t)$ can be written as follows:
\begin{equation}\label{eq:decomp}
\mathbf{U}(\mathbf{x},t) = \Re \left(\sum_{n,m,s}
a_{nms}(t)\,\mathbf{v}_{n m s}(\mathbf{x})\,e^{i\omega_{n m s}
t}\right)
\end{equation}
where $\mathbf{v}_{n m s}(\mathbf{x})$ is the (complex) spatial
structure of the $[n,m,s]$ mode and $\omega_{n m s}$ its
pulsation, which lies in the range $[0, 2\Omega]$ allowed for
inertial waves. Here, $\Re(X)$ stands for the real part of the
complex number $X$. With notations similar to that of
Maas\cite{maas2003} and Bewley {\it et al.},\cite{bewley2007} we
label here the modes using two integer indices, $n$ and $m$, and a
sign, $s=\pm$. The first index $n$ is the normalized vertical
wavenumber such that the horizontal (resp. vertical) velocity
component has $n$ (resp. $n-1$) nodes in the vertical direction.
The horizontal structure of a given mode of vertical index $n$ is
characterized by the second index $m$. Larger values of $m$
essentially correspond to finer structures in the horizontal
direction, although it is not directly related to the number of
nodes as for the vertical index $n$. Finally, the sign $s$ refers
to the symmetry of the mode with respect to the rotation axis:
$s=+$ for a symmetric mode and $s=-$ for an antisymmetric mode.
The pulsations $\omega_{nms}$ are increasing functions of $n$ and,
at fixed $n$ and $s$, decreasing functions of $m$. In the absence
of coupling with other modes or with turbulence, the amplitude of
each mode, $a_{nms}(t)$, is expected to be a decreasing  function
of time because of viscous damping.

The modes present in the mean flow are identified from the
temporal Fourier transform of the ensemble-averaged velocity field
$\mathbf{U}$ at each position in the ($x,z$) measurement plane,
\begin{equation}\label{eq:fft}
\hat{\mathbf{U}}_\sigma (x,z) = \int_{T_{min}}^{T_{max}}
\mathbf{U}(x,z,t)\,e^{-i\sigma t}\,dt.
\end{equation}
The temporal bounds $T_{min}$ and $T_{max}$ have been chosen
equal to 1 and 15 rotation periods, respectively.
From this, we define the temporal energy
spectrum of the mean flow, spatially  averaged over the $(x,z)$
plane, as:
\begin{equation}\label{eq:en}
E_{mean}(\sigma) =  \frac{1}{2\pi} \langle
|\hat{\mathbf{U}}_\sigma (x,z) |^2 \rangle_{x,z}.
\end{equation}

\begin{figure}
\centerline{\includegraphics[width=7cm]{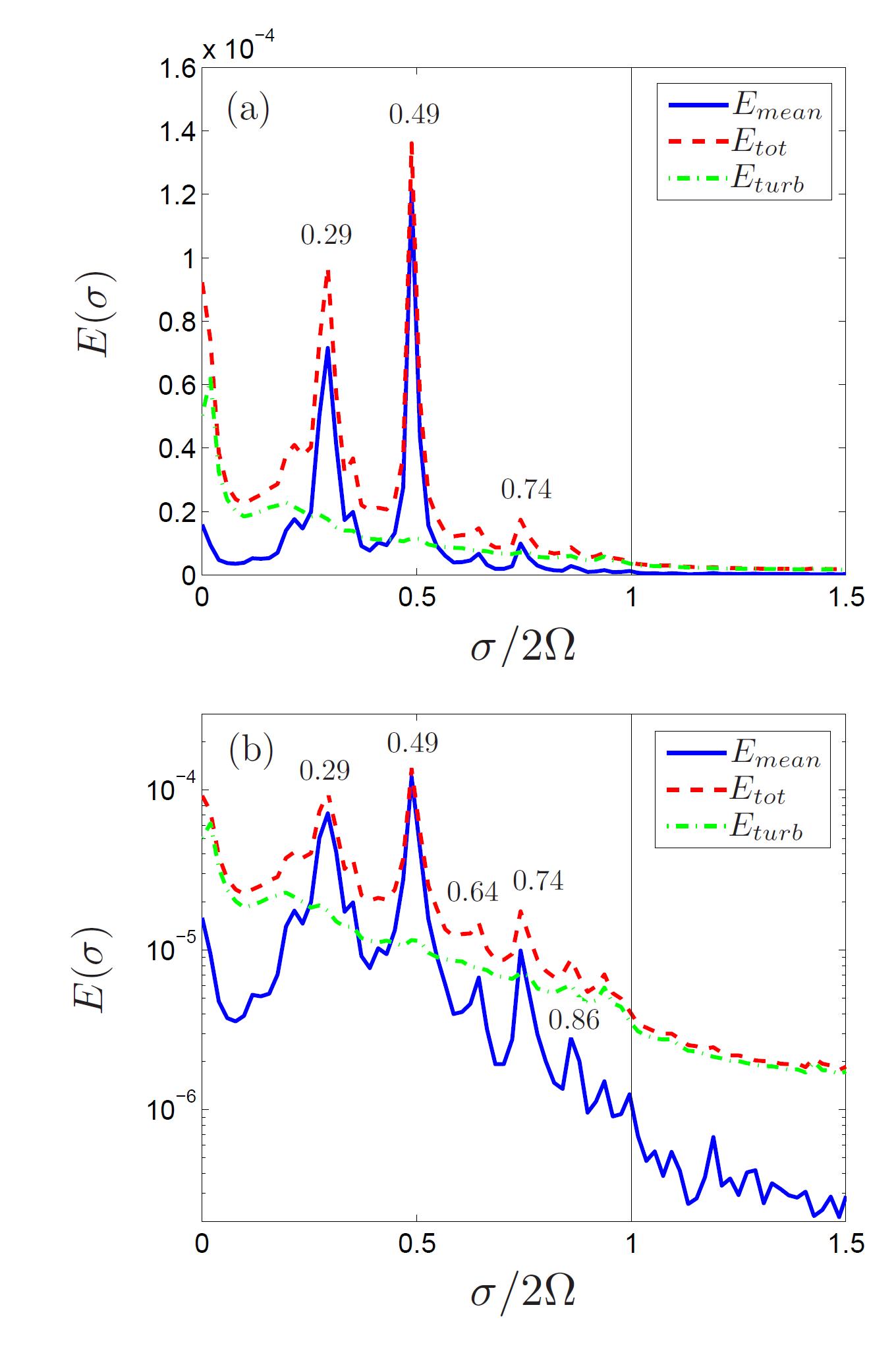}}
\caption{(Color online) Temporal energy spectrum of the total
(dashed), mean (continuous) and turbulent (dashed-dotted)
component of the flow as a function of $\sigma /2\Omega$ computed
from 40 decay realizations performed at
$\Omega=0.84$~rad~s$^{-1}$, with (a) linear and (b) logarithmic
$y$-axis. Inertial modes can develop for pulsations
$\sigma<2\Omega$. The modes corresponding to the peak frequencies
are given in Table~\ref{tab:1}. }\label{fig:spectreenergy}
\end{figure}

In Fig.~\ref{fig:spectreenergy}, the spectrum $E_{mean}(\sigma)$
clearly shows a series of peaks, whose values are listed in
Table~\ref{tab:1}. These peak frequencies are in excellent
agreement with some of the numerical eigenfrequencies computed (as
in Ref.\cite{maas2003}) for our experimental aspect ratio $L/H
\simeq 1.22$. This shows that the translation of the grid, because
of its specific drag profile, selects only a small set of specific
inertial modes among the dense spectrum of modes evidenced in
Ref.\cite{maas2003} In the next subsection, the spatial structures
of the observed modes will be confirmed to match the theoretical
structures predicted for the $[n,m,s]$ modes of Table
~\ref{tab:1}. Those selected modes are among the lowest order
modes ($n = 1$ to $3$), probably because their simpler structure
better matches the spatial features of the ensemble-averaged flow
at early times, i.e. before it is affected by rotation, and also
because higher order modes decay faster by viscous damping.

\begin{table}
\begin{tabular}{ccc}
\hline \hline
 mode & \multicolumn{2}{c}{ $\sigma/2\Omega$ } \\
 \cline{2-3}
$~~~~~~[n,m,s]~~~~~~$ & ~~~~~~num.~~~~~~  & ~~~~~~exp.~~~~~~ \\
\hline
$[1,4,+]$ & 0.2992 & 0.29 \\
$[1,1,+]$ & 0.4890 & 0.49 \\
$[1,1,-]$ & 0.6328 & 0.64 \\
$[2,1,+]$ & 0.7429 & 0.74 \\
$[3,1,+]$ & 0.8557 & 0.86 \\
\hline \hline
\end{tabular}
\caption{Numerical values of normalized frequencies
$\sigma/2\Omega$ for different modes of order $[n,m,s]$ (where $n$
is the vertical wavenumber, $m$ characterizes the horizontal
structure, and $s$ is the symmetry of the mode) compared to the
experimental peaks in Fig.~\ref{fig:spectreenergy}. The
uncertainty of the experimental values is $\pm 0.01$. The
numerical values are computed for an aspect ratio identical the
experimental one $L/H = 1.22$. \label{tab:1}}
\end{table}

In order to check whether the inertial modes are present only in
the ensemble-averaged flow, we also introduce the temporal Fourier
transform of the total, $\hat{\mathbf{u}}_\sigma (x,z)$,  and of
the turbulent velocity field, $\mathbf{\hat{u}^\prime}_\sigma
(x,z)$, similarly to Eq.~(\ref{eq:fft}), from which we can define
the corresponding energy spectra:
\begin{eqnarray}\label{eq:en2}
E_{tot}(\sigma) &=&  \frac{1}{2\pi} \langle \overline{
|\hat{\mathbf{u}}_\sigma|^2 } \rangle, \\
E_{turb}(\sigma) &=&  \frac{1}{2\pi} \langle \overline{
|\mathbf{\hat{u^\prime}}_\sigma|^2 } \rangle.
\end{eqnarray}
As for the kinetic energies, the energy spectra are additive,
$E_{tot}(\sigma) = E_{mean}(\sigma) + E_{turb}(\sigma)$. For a
given total spectrum $E_{tot} (\sigma)$, the two spectra
$E_{mean}$ and $E_{turb}$ allow us to distinguish between two
kinds of inertial modes: (i) reproducible (i.e. phase coherent)
modes, characterized by peaks in $E_{mean}$ but not in $E_{turb}$;
(ii) non-reproducible (i.e. with random phase) modes,
characterized by peaks in  $E_{turb}$ but not in $E_{mean}$.
Interestingly, Fig.~\ref{fig:spectreenergy} shows the complete
absence of peaks in the turbulent spectrum $E_{turb}(\sigma)$,
indicating that all the inertial modes present in our system are
reproducible, and therefore not turbulent. This confirms that
inertial modes in this system are more likely excited by
reproducible flow features induced by the grid translation. This
also suggests that the initial turbulence right after the grid
translation is effectively not affected by the rotation, as could
be expected from its relatively large grid Rossby number.
Therefore, this initial turbulence is expected to be similar to
classical grid turbulence in non-rotating systems.

Finally, it is important to note that more than 97\% of the energy
of the mean flow lies in the range of pulsations $[0, 2\Omega]$,
confirming that almost all its energy is stored in inertial modes.
This result suggests that the ensemble averaged flow is correctly
converged, so that the spectrum of the mean flow is almost not
contaminated by the spectrum of residual turbulent fluctuations
which spans over larger pulsations. On the contrary, we see that
there is a significant amount of energy for $\sigma>2\Omega$ in
the turbulent spectrum, which actually corresponds to the rapid
small scales of usual 3D turbulence which are not directly
affected by the rotation.

\subsection{Spatial structure of the inertial modes}

By performing a band-pass filtering of the mean velocity field at
the frequencies of the peaks identified in $E_{mean}(\sigma)$, it
is possible to extract the spatial structure of each inertial
mode. However, this band-pass filtering must take into account the
decay of the amplitude $a_{nms}(t)$ of each mode, which is
actually reflected in the width of the peaks. Inserting
Eq.~(\ref{eq:decomp}) into Eq.~(\ref{eq:fft}) yields
\begin{equation}
\label{eq:fft2} \hat{\mathbf{U}}_\sigma(\mathbf{x}) = \sum_{n,m,s}
\int_{T_{min}}^{T_{max}}
a_{nms}(t)\,\mathbf{v}_{nms}(\mathbf{x})\,e^{i(\omega_{nms}
-\sigma)t}dt.
\end{equation}
If the selected frequency $\sigma$ coincides with one of the peak
frequency $\omega_{nms}$ identified in the spectrum, the extracted
field is then equal to the spatial structure of the corresponding
mode $[n,m,s]$ weighted by its average amplitude,
\begin{equation}
\hat{\mathbf{U}}_{\omega_{nms}} (x,z) = \mathbf{v}_{nms}(x,z)
\int_{T_{min}}^{T_{max}} a_{nms}(t) dt.
\end{equation}

\begin{figure}
\centerline{\includegraphics[width=8cm]{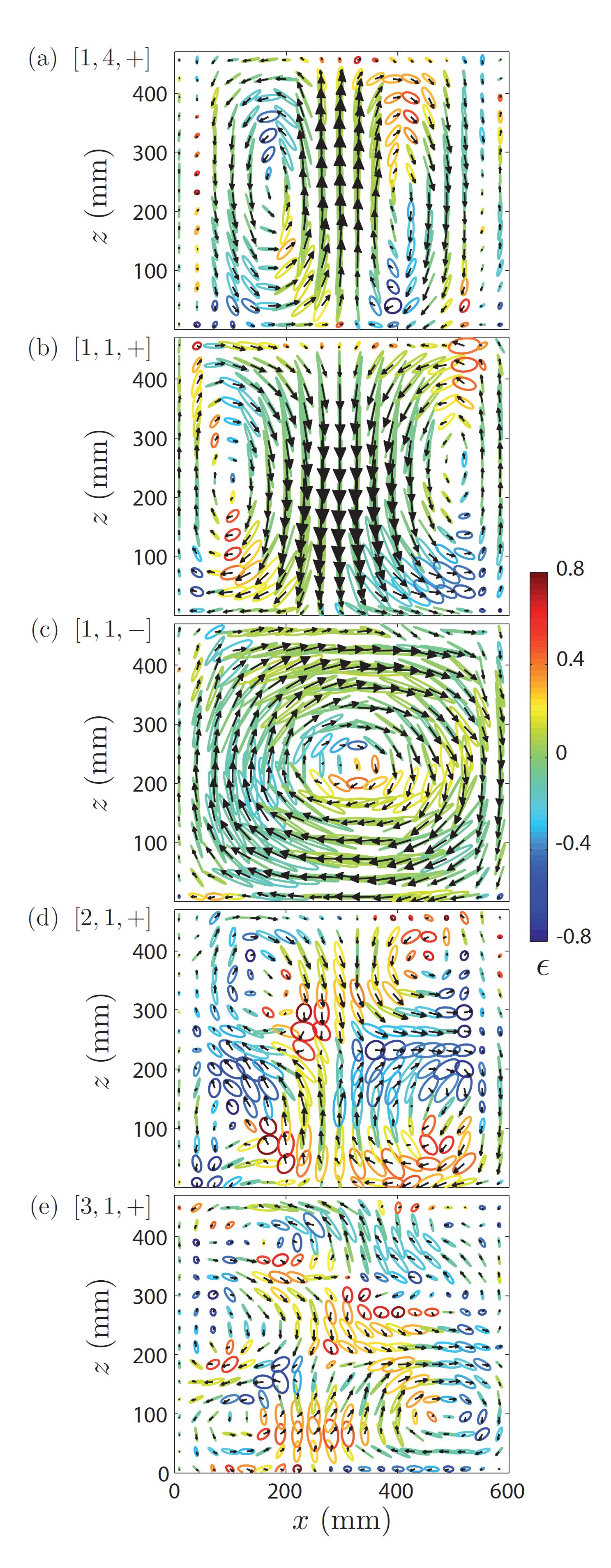}}
\caption{(Color) Spatial structure of the 5 dominating inertial
modes listed in Table~\ref{tab:1}, extracted by band-pass
filtering of the ensemble-averaged fields. The ellipses show the
velocity orbit, and the arrows illustrate the velocity field at a
given arbitrary phase of the oscillation. The color of the
ellipses traces the ellipticity $-1<\epsilon<1$ (see the text for
details). (a) $[1,4,+]$; (b) $[1,1,+]$; (c) $[1,1,-]$; (d)
$[2,1,+]$; (e) $[3,1,+]$. Resolution of the fields has been
reduced by a factor 6 for a better visibility.}
\label{fig:structspat}
\end{figure}

The resulting spatial structure
$\hat{\mathbf{U}}_{\omega_{nms}}(x,z)$ is a complex vector field,
whose real and imaginary parts encode the geometric features
(amplitude, ellipticity, orientation and phase) of the elliptic
orbit described by the velocity vector at each location during one
period. Those orbits can be simply reconstructed by plotting the
trace of the real oscillating vector field,
$\Re(\hat{\mathbf{U}}_{\omega_{nms}} e^{i \omega_{nms} t})$, for
$t \in [0, 2\pi / \omega_{nms}]$. The resulting plots for the 5
dominating inertial modes listed in Table~\ref{tab:1} are shown in
Fig.~\ref{fig:structspat}. The velocity field, taken for an
arbitrary phase of the oscillation, is also illustrated by the
vector arrows. The color of the ellipses corresponds to the local
ellipticity $-1<\epsilon<1$, where positive ellipticity
corresponds to an anticlockwise rotation.

The geometrical features of these fields provide a good
qualitative support for the identification of the frequencies
given in Table~\ref{tab:1}. In particular, the modes of larger
energy are of vertical wavenumber $n=1$
[Figs.~\ref{fig:structspat}(a-c)], i.e. they show one cell in the
vertical direction. Higher order vertical modes, $n=2$ and $n=3$,
are also found, with much weaker energy. The irregular shape of
the last mode, $n=3$, is probably due to its very low energy
level, and to a possible mixing with other modes of similar
frequency. Except for the antisymmetric mode $[1,1,-]$ at
$\sigma/2\Omega=0.64$ [Fig.~\ref{fig:structspat}(c)], all the
excited modes are symmetric ($s=+$) with respect to the rotation
axis. This weak antisymmetric mode is probably excited from a
spontaneous symmetry breaking of the flow, or from a slight
residual left-right asymmetry of the grid.

The horizontal structure of the modes cannot be inferred from
these fields in the vertical plane. However, the visualizations
provided in Ref.\cite{maas2003} indicate that, in spite of the
square shape of the container, the modes are approximately
axisymmetric, at least at a moderate distance from the rotation
axis.

\begin{figure}
\centerline{\includegraphics[width=7.5cm]{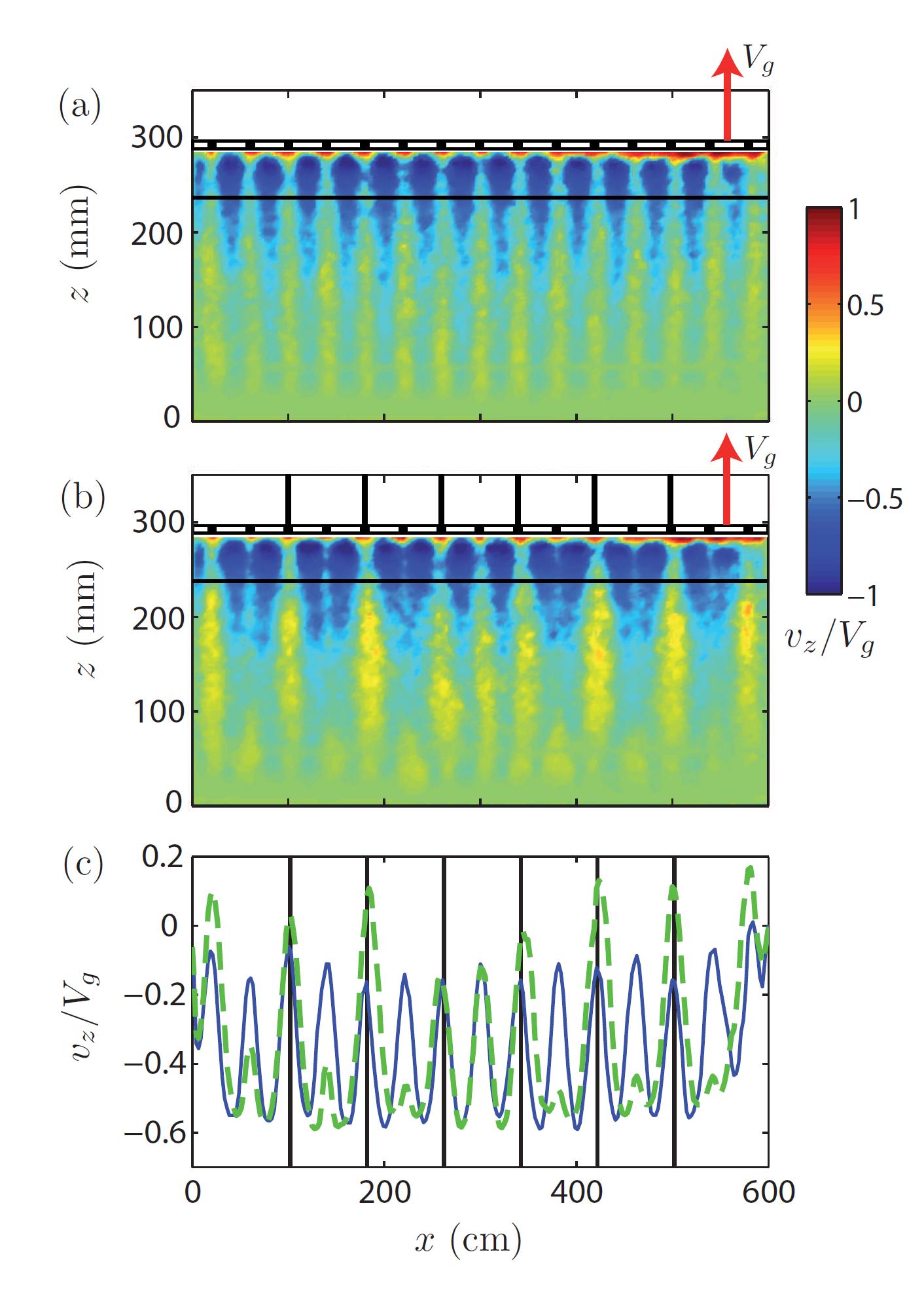}}
\caption{(Color) Vertical velocity field in the wake of the grid.
(a) Simple grid configuration. (b) Modified configuration with the
inner walls attached to the grid, represented by the 6 vertical
thick lines. The grid is towed from the bottom, and is at the
height $z=290$~mm in these snapshots. (c) Horizontal profile of
the vertical velocity, at a distance $\delta z=5$~cm below the
grid (indicated by the horizontal lines in a and b). Continuous
line: simple grid configuration. Dashed line: modified
configuration with inner walls. The vertical black lines show the
locations of the vertical inner walls.}\label{fig:vitgrille}
\end{figure}

The selection of the excited inertial modes among all the possible
resonant ones may originate from a residual inhomogeneity of the
jet velocities through each mesh of the grid. Such inhomogeneity
is not necessarily associated to a defect in the grid, but could
arise from the influence of the side walls of the tank or from a
spontaneous symmetry breaking of the flow. In particular, the
spatial structure of the two dominant modes (at $\sigma / 2 \Omega
= 0.29$ and $0.49$) suggests that larger (or smaller) velocities
are produced near the center of the grid, initiating a marked
vertical columnar oscillation.

In order to check the possible inhomogeneity of the flow produced
by the grid, we have measured the velocity profile just behind the
grid during its translation (Fig. \ref{fig:vitgrille}a, and the
continuous curve in Fig. \ref{fig:vitgrille}c). Slightly weaker
jets are indeed encountered near the borders of the grid, probably
originating from the friction with the sidewalls. The resulting
rounded average profile could be responsible for the excitation of
the dominant symmetric modes.

\section{Modified configuration with inner walls}
\label{sec:giw}

\subsection{Comparison with a previous configuration}

Although the excitation mechanism of inertial modes in a rotating
grid turbulence experiment is interesting in itself, generating
such turbulence without those reproducible modes in a confined
geometry is also highly desirable. This is motivated by the fact
that most theoretical and numerical works on this subject have
been performed in purely homogeneous rotating turbulence, without
inertial modes (see, e.g., Sagaut and Cambon\cite{Cambon} and
references therein). In this section, we characterize the flow
generated by a modified grid configuration, which turns out to be
almost free of inertial modes.

In a similar rotating turbulence experiment, in order to avoid a
large scale recirculation flow triggered by the grid translation,
Staplehurst \textit{et al.}\cite{staplehurst2008} have designed an
original setup in which a parallelepipedic inner tank, consisting
in 4 vertical sidewalls without top and bottom walls, was attached
on the upper side of the grid. This inner tank, of 35~cm side, was
translated with the grid from the {\it top to the bottom}, inside
the primary 45~cm wide square container filled with water. In this
situation, the inner sidewalls are therefore located {\it
downstream} the grid, and once the grid is lowered, the effective
flow volume is reduced to the size of the inner tank. With this
additional inner tank, these authors actually succeeded to reduce
the recirculation flow by a factor 5. However, only the
non-rotating case was described, and the possible presence of
inertial modes with rotation has not been investigated.

\begin{figure}
\centerline{\includegraphics[width=6.5cm]{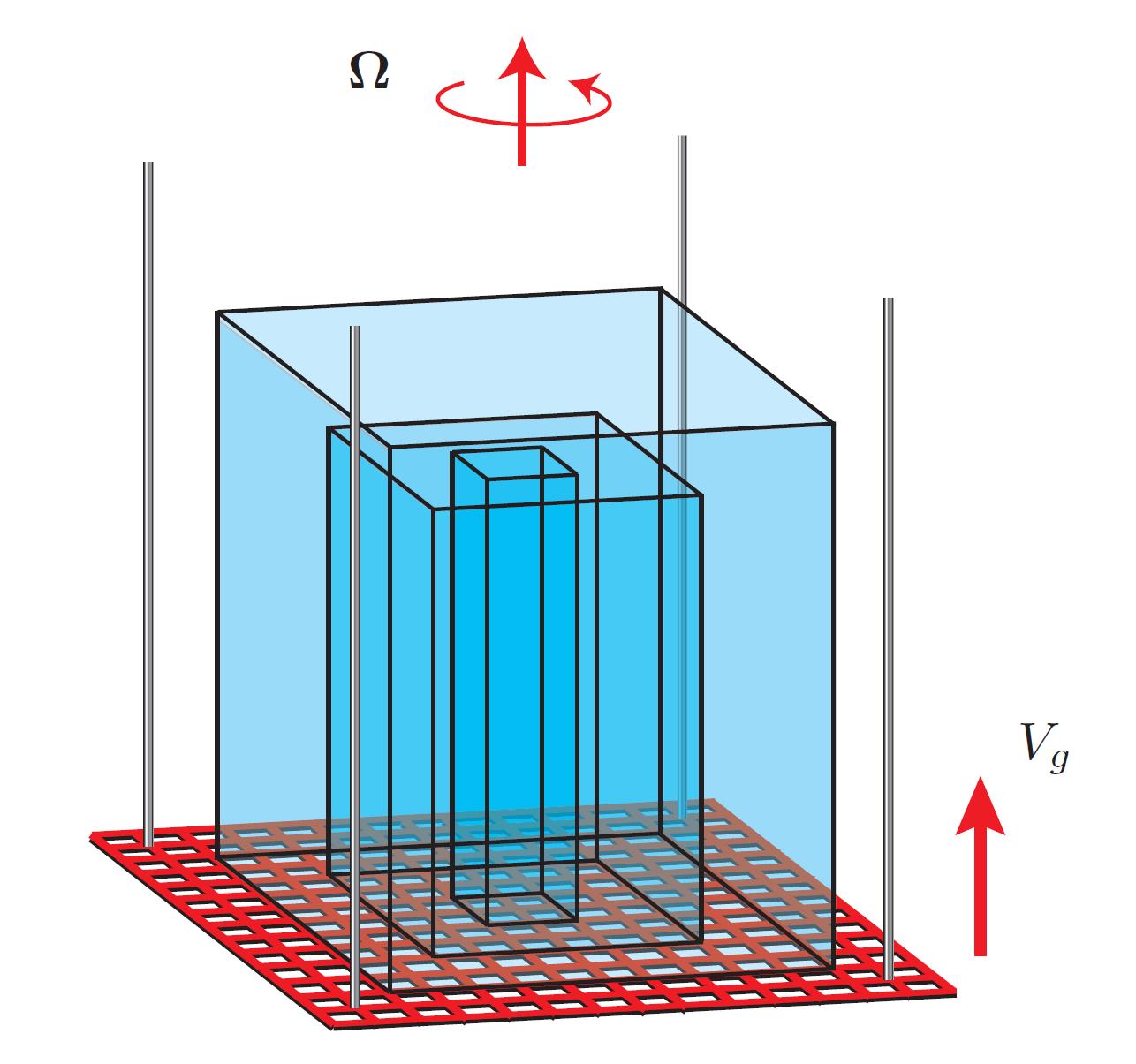}}
\caption{(Color online) Schematic view of the modified grid
configuration (the outer water tank is not shown; see
Fig.~\ref{fig:setup}). Three inner tanks are mounted on the grid,
and turbulence is generated by raising the set grid+tanks. Each
inner tank consists in 4 vertical sidewalls, without top and
bottom walls.} \label{fig:setup2}
\end{figure}

The purpose of the inner tank in the previous downstream
configuration was to block the transverse motions in the wake of
the grid. Here, in order to force equally distributed jets through
the grid, while keeping the full volume for the working fluid, we
propose to modify the solution of Staplehurst \textit{et
al.},\cite{staplehurst2008} by adding inner walls located {\it
upstream} the grid. In our configuration, the inner walls are
still attached on the upper side of the grid, but the grid here is
lifted from the bottom to the top. As a consequence,  the working
fluid volume in the wake of the grid is the same as in the simple
grid configuration, so that direct comparisons of the flow
structure can be performed between the two configurations. Several
configurations have been tested, with 1, 2 and 3 inner tanks. Best
results have been obtained with 3 inner tanks, and only this
configuration is described in the following.

\subsection{Modified experimental setup}

The new setup is shown in Fig.~\ref{fig:setup2}. It consists in
three parallelepipedic PVC tanks, of respective height 50, 40 and
40~cm and width 40, 24 and 8~cm. Each tank consists in 4 vertical
sidewalls, without top and bottom walls. The 3 tanks are fit
together co-axially, so that two neighbour sidewalls are separated
by two grid meshes, and attached on the upper side of the grid.
All other parameters, in particular the grid velocity and Rossby
number, are the same as for the simple grid configuration. Since
our grid is translated from the bottom to the top, here the inner
tanks are {\it upstream} the grid, so that the decaying flow is
not confined inside the smallest tank.

The presence of the inner tanks implies that no horizontal cover
could be placed to define the upper boundary of the flow,
contrarily to the ``simple'' grid configuration. However, we can
consider here that an effective upper boundary is approximately
defined by the grid itself, locked at a height of 46~cm after its
translation. Indeed, we have checked that the ``simple'' grid
configuration without upper rigid boundary also produces inertial
modes, which are as intense as with the rigid boundary. Moreover,
here, the inner walls efficiently block the flow in the region
between the grid and the free surface.

The velocity field and profiles in this modified configuration are
shown in Figs. \ref{fig:vitgrille} (b) and (c), compared to that
obtained with the simple grid configuration [Figs.
\ref{fig:vitgrille}(a) and (c)]. The main difference is that, now,
a stronger upper flow is present in the wakes of those grid bars
supporting the sidewalls, whereas the upper flow in the wake of
the other grid bars is significantly reduced, and actually even
difficult to detect with our PIV resolution. As a consequence, two
neighbor jets originating from the same ``passage'' between two
sidewalls tend to gradually merge. At this point, comparing the
velocity profiles of the two configurations, with and without the
inner tanks, it is not clear whether changes in the energy budget
between turbulence and mean flow may arise from the observed
differences.

\subsection{Turbulence generated with the grid + inner tanks configuration}

\begin{figure}
\centerline{\includegraphics[width=8cm]{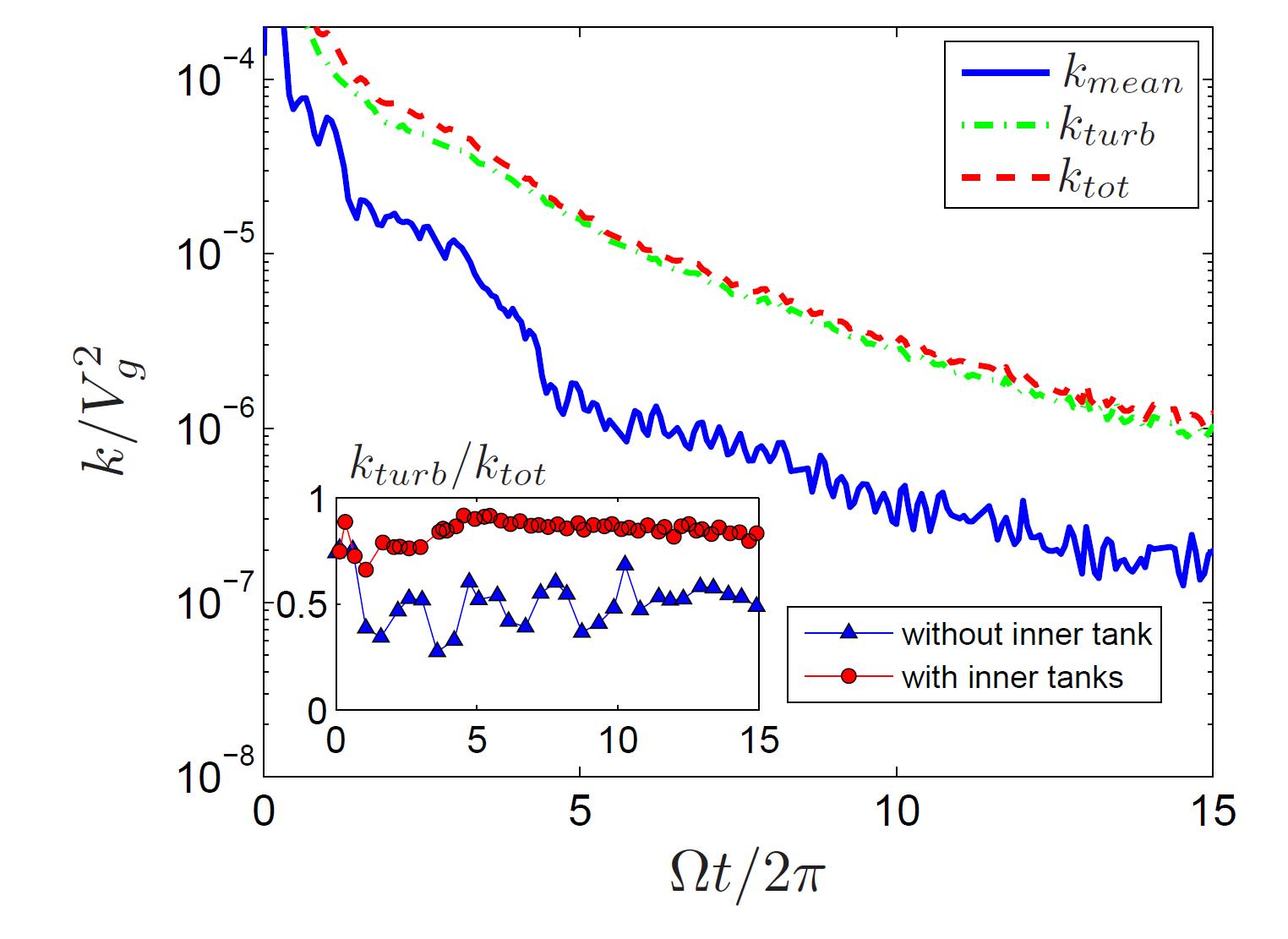}}
\caption{(Color online) Total (dashed), mean (continuous) and
turbulent (dashed-dotted) kinetic energies as a function of
reduced time $\Omega t/2\pi$ from 40 decay realizations performed
at $\Omega=0.84$~rad~s$^{-1}$ with the modified configuration with
inner tanks. Inset: ratio of turbulent to total kinetic energy as
a function of reduced time $\Omega t/2\pi$ with (points) and
without (triangles) inner tanks.}\label{fig:decay2}
\end{figure}

The decay of the kinetic energies of the total, mean and turbulent
components of the flow as a function of time, from ensemble
averages over 40 decay realizations with the modified
configuration, is shown in Fig.~\ref{fig:decay2}. In comparison
with  the simple grid configuration (Fig.~\ref{fig:decay1}), the
oscillations due to the inertial modes are strongly reduced, both
for the mean and the total kinetic energies. We also see that,
contrary to the simple grid configuration, the turbulent kinetic
energy is now significantly larger than the one of the mean flow.
This can be better seen in the inset of Fig. \ref{fig:decay2},
showing that turbulence contains, after a transient of about one
tank rotation, approximately $85 \pm 5\%$ of the total kinetic
energy, a value much larger than the 50\% obtained with the simple
grid configuration.

\begin{figure}
\centerline{\includegraphics[width=7cm]{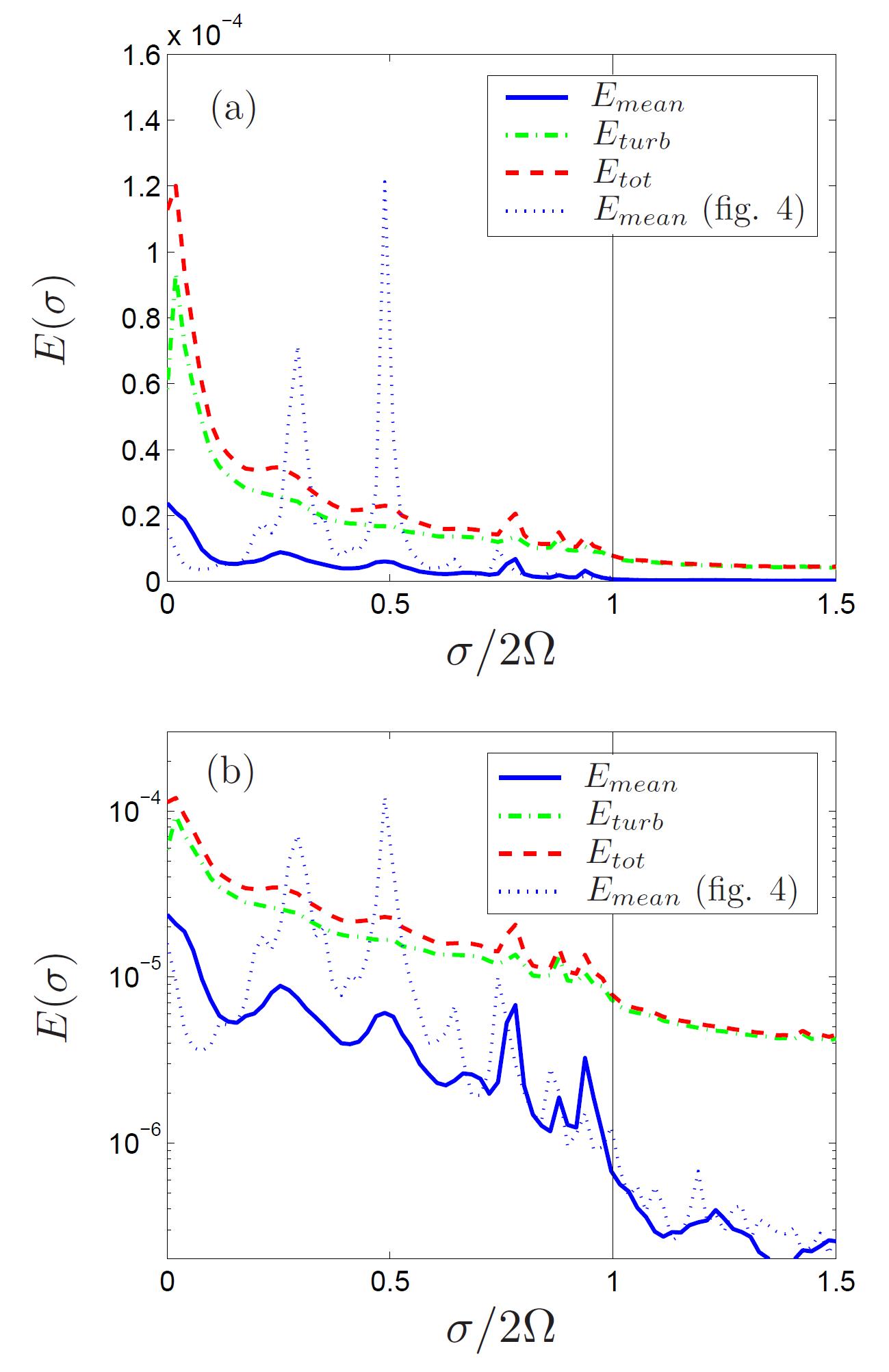}}
\caption{(Color online) Temporal energy spectrum of the total
(dashed), mean (continuous) and turbulent (dashed-dotted)
component of the flow as a function of $\sigma /2\Omega$ computed
from 40 decay realizations performed at $\Omega=0.84$~rad~s$^{-1}$
with inner tanks, with (a) linear and (b) logarithmic $y$-axis.
The dotted line reproduces spectrum of the mean flow for
comparison (Fig. \ref{fig:spectreenergy}), obtained with the
simple grid configuration.} \label{fig:spectreenergy2}
\end{figure}

In Fig. \ref{fig:spectreenergy2}, we superimpose the temporal
energy spectrum of the mean flows obtained with (continuous line)
and without (dotted line) the inner tanks. We clearly see that,
with the inner tanks, the two dominant peaks (at $\sigma/2\Omega =
0.29$ and 0.49) have been reduced by more than a factor of 10. The
reduction of the other peaks, in particular the $n=2$ mode at
$\sigma/2\Omega \simeq 0.74$, is less pronounced. The decrease of
the two dominant inertial modes essentially explains the
significant drop of the kinetic energy of the mean flow with
respect to turbulence shown in Fig. \ref{fig:decay2}. We can also
note slight frequency shifts between the two configurations,
probably originating from the not perfectly identical boundary
conditions at the top of the container, and hence from slightly
different effective aspect ratios.

Interestingly, inertial modes are not only absent from the mean
flow spectrum, $E_{mean}(\sigma)$ but also from the turbulent
spectrum $E_{turb}(\sigma)$. This indicates that their
disappearance in the modified grid configuration cannot be
attributed to  a change in the nature of the excited inertial
modes, from reproducible to non reproducible, but to a true
disappearance.

The differences in the energy decays and spectra show that the
addition of the inner tanks to the grid leads to a much more
efficient transfer of the initial mean flow energy into
turbulence, bypassing the generation of inertial modes. However,
the origin of the changes between the two configurations is not
clear. In particular, it is not evident if, and how, the
differences in the jet profiles in Fig.~\ref{fig:vitgrille}(c) may
explain the preferential energy transfer from the mean flow to the
turbulence in the modified configuration.  These results
illustrate the sensitivity of the inertial modes production to
slight changes in the grid geometry, at least for the grid Rossby
number considered here, $Ro_g = 10.4$.

\section{Discussion and conclusion}

In this article, we characterize in details the flow generated
when a simple grid is rapidly translated in a rotating and
confined volume of fluid. We show that, in addition to a
non-reproducible turbulent flow, the grid translation initiates a
reproducible mean flow composed of resonant inertial modes, which
contains a significant amount of the total kinetic energy. This
observation agrees with the recent experiments of Bewley {\it et
al.}\cite{bewley2007} in a similar geometry. In addition to
quantitative comparisons of the mode frequencies with the
numerical predictions of Maas\cite{maas2003}, we also provide here
good qualitative comparisons for the spatial structure of these
modes.

These results suggest that the turbulent (i.e. the non
reproducible) component of the flow generated in this
configuration cannot be considered as freely decaying. Indeed,
energy transfers between the inertial modes and the turbulence may
exist, so that the energy initially stored in the inertial modes
may continuously feed the turbulence. However, measuring this
transfer would require to compute the turbulent Reynolds stress
tensor and its coupling with the ensemble-averaged flow. This
cannot be performed from the PIV data presented here, which are
limited to two velocity components in a single plane.

The ability of a translated grid to generate or not reproducible
inertial waves or modes depends on the grid Rossby number, $Ro_g =
V_g / 2\Omega M$, and also on the geometrical details of the grid
configuration, in particular its solidity (ratio of solid to total
area). Indeed, for a given grid geometry, a low Rossby number
seems more likely to excite inertial modes, because the
reproducible wake pattern of the grid, or the induced reproducible
recirculation flow, may directly force inertial modes with a phase
coherence. On the other hand, we could speculate that, for much
larger Rossby numbers (and assuming that the Reynolds number is
very large too), the reproducible wake pattern has more time to
loose its coherence before being affected by rotation, so its
energy may be more efficiently transferred to turbulence than
reproducible inertial modes. Even if inertial waves are generated
from these decaying incoherent motions, their random phase should
cancel them out in the ensemble average, so inertial modes should
not be found in the mean flow.

In the experiments of Bewley {\it et al.},\cite{bewley2007} the
moderate grid Rossby number $Ro_g$ of 5.5 is found to produce
strong inertial oscillations. In the experiments of Staplehurst
{\it et al.},\cite{staplehurst2008} $Ro_g$ lies in the range
$1-3.2$, so that significant inertial modes are also expected,
although they are not described by these authors. Finally, in the
present experiment, for a slightly larger grid Rossby number of
10.4, the production of reproducible inertial modes is found to be
very sensitive to slight details in the geometry of the grid.
Indeed, by adding a set of inner tanks to the grid, we show that
it is possible to significantly reduce the production of inertial
modes, although the mechanism responsible for this preferred
energy transfer towards turbulence instead of modes is not
elucidated in the present work. This sensitivity of the flow
evidenced for this particular $Ro_g$ is consistent with the
results of Morize and Moisy,\cite{morize2006} covering a range of
$Ro_g$ between 2 and 65, in which a smooth transition around $Ro_g
\simeq 10$ is found in the decay law of the turbulent energy,
suggesting a change in the energy transfer between inertial modes
and turbulence.

From those results, one may conclude that, providing the grid
Rossby number and the geometrical features of the grid are
carefully selected, achieving a nearly ``pure'' rotating
turbulence, free of reproducible inertial modes, is actually
possible in a confined geometry. This system is therefore suitable
to explore experimentally the influence of the rotation on freely
decaying turbulence.

\acknowledgments

We acknowledge C. Morize and M. Rabaud for discussions about the
manuscript, and A. Aubertin, L. Auffray, C. Borget, G.-J. Michon
and R. Pidoux for experimental help. The rotating platform
``Gyroflow'' was funded by the ANR (grant no. 06-BLAN-0363-01
``HiSpeedPIV''), and the ``Triangle de la Physique''.

\end{document}